\documentclass{PoS}
\usepackage{subcaption}
\usepackage{amssymb}
\usepackage{amsmath}

\title{Blazar Alerts with the HAWC Online Flare Monitor }

\ShortTitle{Blazar Alerts with HAWC}

\author{\speaker{Thomas Weisgarber}$^a$ and Ian G. Wisher$^a$ for the HAWC Collaboration$^b$ \\
        \llap{$^a$}Wisconsin IceCube Particle Astrophysics Center (WIPAC) and Department of Physics, University of Wisconsin---Madison, Madison, WI, USA \\
        \llap{$^b$}For a complete author list, see \href{http://www.hawc-observatory.org/collaboration/icrc2015.php}{www.hawc-observatory.org/collaboration/icrc2015.php}. \\
        Email: \email{weisgarber@physics.wisc.edu}, \email{iwisher@icecube.wisc.edu}}

\abstract{
The High Altitude Water Cherenkov (HAWC) Observatory monitors the gamma-ray sky in the 100 GeV to 100 TeV energy range with $>95\%$ uptime and unprecedented sensitivity for a survey instrument.
The HAWC Collaboration has implemented an online flare monitor that detects episodes of rapid flaring activity from extragalactic very high energy (VHE) sources in the declination band from -26 to 64 degrees.
This allows timely alerts to be sent to multiwavelength instruments without human intervention.
The preliminary configuration of the online flare monitor achieves sensitivity to flares of at least 1 hour duration that attain an average flux of 10 times that of the Crab Nebula.
While flares of this magnitude are not common, several flares reaching the level of 10 Crab have been observed in the VHE band within the past decade.
With its survey capabilities and high duty cycle, HAWC will expand the observational data set on these particularly extreme flares.
We characterize the sensitivity of the online flare monitor and outline plans for its upcoming deployment.
}

\FullConference{The 34th International Cosmic Ray Conference,\\
		30 July- 6 August, 2015\\
		The Hague, The Netherlands}

\begin{document}

\section{Blazar Flares at Very High Energies}
\label{section:intro}

Blazars comprise a class of active galactic nuclei (AGNs) with relativistic jets oriented along the line of sight to Earth~\cite{1995PASP..107..803U}.
More than 50 objects in this class have been detected in the very high energy (VHE; $E>100$ GeV) band\footnote{See http://tevcat.uchicago.edu for an up to date list.}.
Typical blazar spectra exhibit a characteristic ``two-bump'' feature, with the low energy bump, spanning radio to hard x-rays, being interpreted as synchrotron emission from a population of relativistic electrons in the jet.
The high energy bump, spanning MeV energies and above, is believed to originate either due to a \textit{leptonic} scenario---in which low energy photons are inverse Compton scattered by the leptons responsible for the synchrotron emission---or due to a \textit{hadronic} scenario---in which proton synchrotron emission, pion decay, or other hadronic interactions dominate the high energy emission (see for example~\cite{2013ApJ...768...54B} and references therein).

Blazars are known to be highly variable sources in the VHE energy band, with fluxes varying on time scales ranging from minutes to months and increasing by more than an order of magnitude over their baseline quiescent states~\cite{2007JPhCS..60..318T}.
The study of these flaring states can offer insight into the mechanisms responsible for powering the relativistic jets, including the possibility to discriminate between leptonic and hadronic origins of the VHE emission.
Furthermore, the increase in high energy event counts during flares provides an opportunity to measure the extragalactic background light via the attenuation of primary gamma rays as they propagate through the intergalactic medium~\cite{2013APh....43..241M}, and the timing characteristics of the flares may be used to measure Lorentz invariance violation~\cite{Nellen:2015} or detect the presence of an intergalactic magnetic field~\cite{2009PhRvD..80l3012N,Weisgarber:2013}.
Identifying flaring blazars and obtaining a data set of multiwavelength observations during the flares is therefore an ongoing endeavor of great scientific interest.

Until recently, VHE observations of blazar flares on hour time scales have been conducted exclusively by imaging atmospheric Cherenkov telescopes (IACTs), which observe the Cherenkov radiation from secondary particles in air showers initiated by high energy gamma rays~\cite{1988PhR...160....1W}.
The current generation of IACTs has collected numerous observations of flaring blazars.
Over the past decade, a few particularly extreme flares, reaching more than 10 times the flux from the Crab Nebula (the strongest steady source of VHE gamma rays), have been observed~\cite{2007ApJ...664L..71A,2013ATel.4976....1C}.
However, since IACTs are pointed instruments with limited sky coverage and are likely to continue observing a source found to be in a flaring state, the catalog of VHE flares is biased, rendering the determination of the rate at which blazars flare difficult.

In contrast to IACTs, VHE survey instruments observe a wide field of view that allows them to monitor a large number of sources simultaneously.
The High Altitude Water Cherenkov (HAWC) Observatory, located at $19^\circ$ N latitude at an elevation of 4100 m in Sierra Negra, Mexico, is the most sensitive VHE survey instrument ever constructed.
The detector comprises 300 optically isolated water Cherenkov detectors, each of which contains 4 upward-facing photomultiplier tubes (PMTs) monitoring 200,000 liters of purified water for the Cherenkov radiation produced by secondary particles propagating through the detector volume.
HAWC, described in detail in~\cite{Pretz:2015}, features a field of view of 2 sr, an uptime of $>95\%$, and a substantial improvement in sensitivity over its predecessors~\cite{2012ApJ...750...63A,2015ApJ...798..119B}, enabling it to conduct an unbiased survey of flaring activity in blazars.
This capability will greatly enhance the existing data set of extreme flares, providing new opportunities to study the blazar central engine, as well as to investigate the propagation of VHE gamma rays over cosmic distances.

Due to its location, HAWC achieves maximum sensitivity to sources with declinations close to $19^\circ$.
For sources culminating at an elevation angle of $45^\circ$ (declinations ranging from $-26^\circ$ to $64^\circ$ for HAWC), the sensitivity is reduced by approximately a factor of 10~\cite{2013APh....50...26A} and falls rapidly outside of this range.
For the present investigation, we restrict our attention to sources within this declination range.
We also focus this study on the sensitivity of HAWC to detect flares on hour time scales, with the intention to issue alerts as quickly as possible after such flaring states are identified.
The sensitivity of HAWC to flares on time scales of 1 day or more and initial observations of such flares are presented elsewhere in these proceedings~\cite{Lauer:2015}.

\section{The HAWC Online Flare Monitor}
\label{section:monitor}

The online flare monitor tracks all blazars known to be VHE emitters, as well as a selection of nearby sources ($z<1$) from the 1FHL catalog produced by the Fermi Collaboration~\cite{2013ApJS..209...34A}.
Events triggering the HAWC detector are reconstructed in real time by the HAWC online system, after which they are processed by the online flare monitor.
The events are divided into analysis bins defined in terms of $n_\textnormal{frac}$, the fractional number of PMTs that participate in the event.
The parameter $n_\textnormal{frac}$ serves as a crude proxy for the total energy deposited by the air shower at ground level.
An improved estimator for the energy of the primary particle will be used to construct the analysis bins in the near future.

Data reduction is achieved via a set of selection cuts that maximize sensitivity to the Crab Nebula.
The selection cuts include the distribution of charge recorded by the PMTs and the quality of the fits that determine the shower core and angular position on the sky.
Separate optimal cuts are determined for each analysis bin.
Further information about the cuts is available in these proceedings~\cite{Smith:2015,Greus:2015}.

The determination of the analysis bins for the online flare monitor is different from that for steady sources as presented in~\cite{Greus:2015}.
Because we seek transient signals on hour time scales, the event rate in each analysis bin must be much larger than 1 per hour.
We therefore define analysis bins to achieve roughly equal count rates in each bin, and we optimize the cuts to maximize the sensitivity to the Crab Nebula.
We have tested the online flare monitor with 1, 2, 3, 4, and 8 analysis bins, and we find that the sensitivity is maximized for the 3-bin case.

\section{Flare Identification}
\label{section:flares}

We define flares as periods during which the flux from a monitored source exhibits a significant increase over its quiescent level.
Because the detector response depends on the zenith angle (and therefore changes as the source moves through HAWC's field of view), simply monitoring the rate of events from the source is insufficient to identify flares.
Instead, we search for changes in the signal to background ratio $R\equiv N_\textnormal{on}/\alpha N_\textnormal{off}$, where following~\cite{1983ApJ...272..317L} we define $N_\textnormal{on}$ and $N_\textnormal{off}$ as the on-source and off-source counts and $\alpha$ as the ratio of on-source to off-source exposure.
The off-source counts $N_\textnormal{off}$ are derived via the direct integration method~\cite{2003ApJ...595..803A}, in which the product of the detector rate and the normalized counts in an angular bin of solid angle $\Omega_D$ in local dectector coordinates is integrated over a pre-defined time period, here chosen to be 2 hours.
The value of $N_\textnormal{on}$ is determined from the counts over the same time interval and in an angular bin, centered on the position of the source, with the same solid angle $\Omega_D$, but in equatorial coordinates instead of local detector coordinates.

Our search employs the Bayesian block algorithm~\cite{2013ApJ...764..167S}, which we use to identify the optimum partitioning of the data into blocks within which $R$ is constant.
The edges of the blocks are then taken as triggers indicating the start or end of flaring activity.
For a given set of $N_p$ data points, the algorithm finds the optimum partition in $N_p^2$ time, compared to the brute force method, which scales as $2^{N_p}$.
Optimality is defined by a fitness function $F$ that specifies the fitness of a given block to be represented by a constant value.
The fitness function must depend only on the data within the block, and it must be additive such that the fitness of a set of blocks is the sum of the individual blocks' fitnesses.

The HAWC events are too numerous to permit an unbinned analysis, so we bin the observations into data points of 1 minute duration.
Each data point $i$ consists of a number of on counts, denoted by $n_i$, and a number of off counts, denoted by $m_i$.
Since the detector response changes with time as the source elevation changes, we assume that each 1-minute interval has a separate true Poisson mean characterizing the distribution from which the off count rate $m_i$ is derived.
We denote these mean values by $\mu_i$.
Our hypothesis for each block is that the ratio $R$ is constant, so we can write the true Poisson mean of the distribution from which the on counts are sampled as $\alpha\mu_iR$.
We then follow Section 3 of~\cite{2013ApJ...764..167S} and construct the fitness function for the block as the sum of the log likelihoods for each data point $i$ in the block:
\begin{equation}
\ln L(\vec{n},\vec{m}|R,\vec{\mu})=\sum_i\left[-\alpha\mu_iR+n_i\ln(\alpha\mu_iR)-\mu_i+m_i\ln\mu_i-g(n_i,m_i)\right],
\label{eqn:block_likelihood}
\end{equation}
where $g(n_i,m_i)\equiv\ln\Gamma(n_i+1)+\ln\Gamma(m_i+1)$ and $\Gamma(x+1)=x!$ is the factorial function.
To derive the fitness function, we choose the values of $R$ and $\mu_i$ that maximize the likelihood, obtaining
\begin{equation}
F_b=-N\left[1+\ln\left(1+\frac{M}{N}\right)\right]-M\left[1+\ln\left(1+\frac{N}{M}\right)\right]+\sum_i\left[(n_i+m_i)\ln(n_i+m_i)-g(n_i,m_i)\right].
\label{eqn:block_fitness_bin}
\end{equation}
Here, $N\equiv\sum_in_i$ and $M\equiv\sum_im_i$, and all sums run over the data points in the block.
Since the HAWC data are divided into analysis bins as well as time bins, the index $b$ in Equation~\ref{eqn:block_fitness_bin} indicates that this is the fitness contribution from one analysis bin.
The total block fitness is then given by
\begin{equation}
F=\sum_bF_b.
\label{eqn:block_fitness}
\end{equation}

The final input to the Bayesian block algorithm is a probability $p$ that accounts for our prior belief in the number of blocks.
Specifically, we assume that it is $p$ times less likely for the data to be described by $N_B+1$ blocks than by $N_B$ blocks, for $N_B\in\{1,2,3,...\}$.
The value of $p$ is then tuned to set a predetermined false positive rate that depends on the monitored source class.
For this study, we fix the prior to obtain a false positive rate of $10^{-5}$ events per minute per monitored source.
This requires different priors to be set for different numbers of analysis bins.
For example, the prior in the case of 1 analysis bin is $p_1\approx10^{-5}$, while the prior for 8 analysis bins is $p_8\approx10^{-9}$.

\section{Flare Monitor Sensitivity}
\label{section:sensitivity}

\begin{figure}
 \begin{subfigure}[t!]{0.5\textwidth}
  \includegraphics[width=\textwidth]{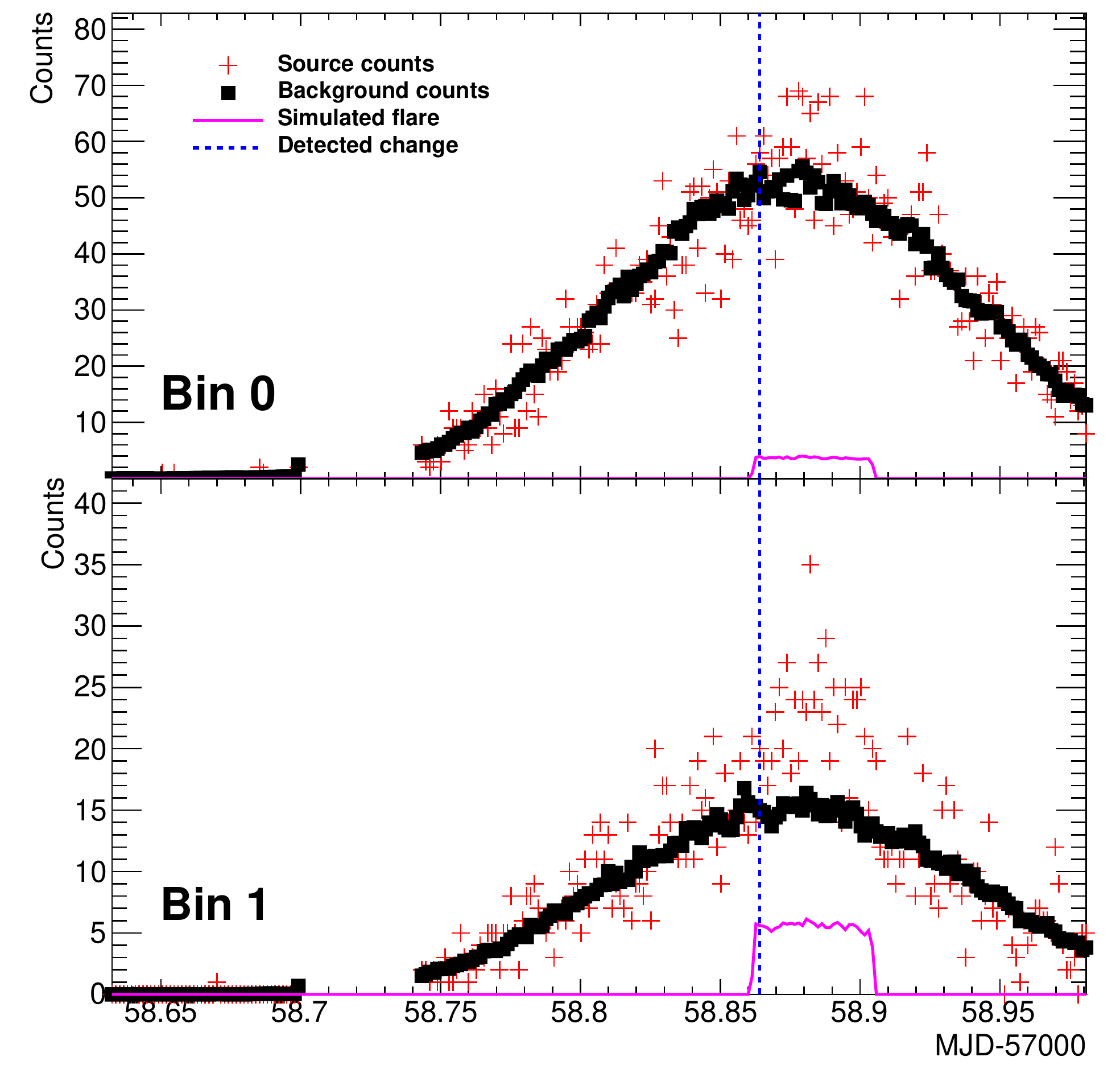}
  \caption{}
  \label{fig:monitor_example}
 \end{subfigure}
 \begin{subfigure}[t!]{0.5\textwidth}
  \includegraphics[width=\textwidth]{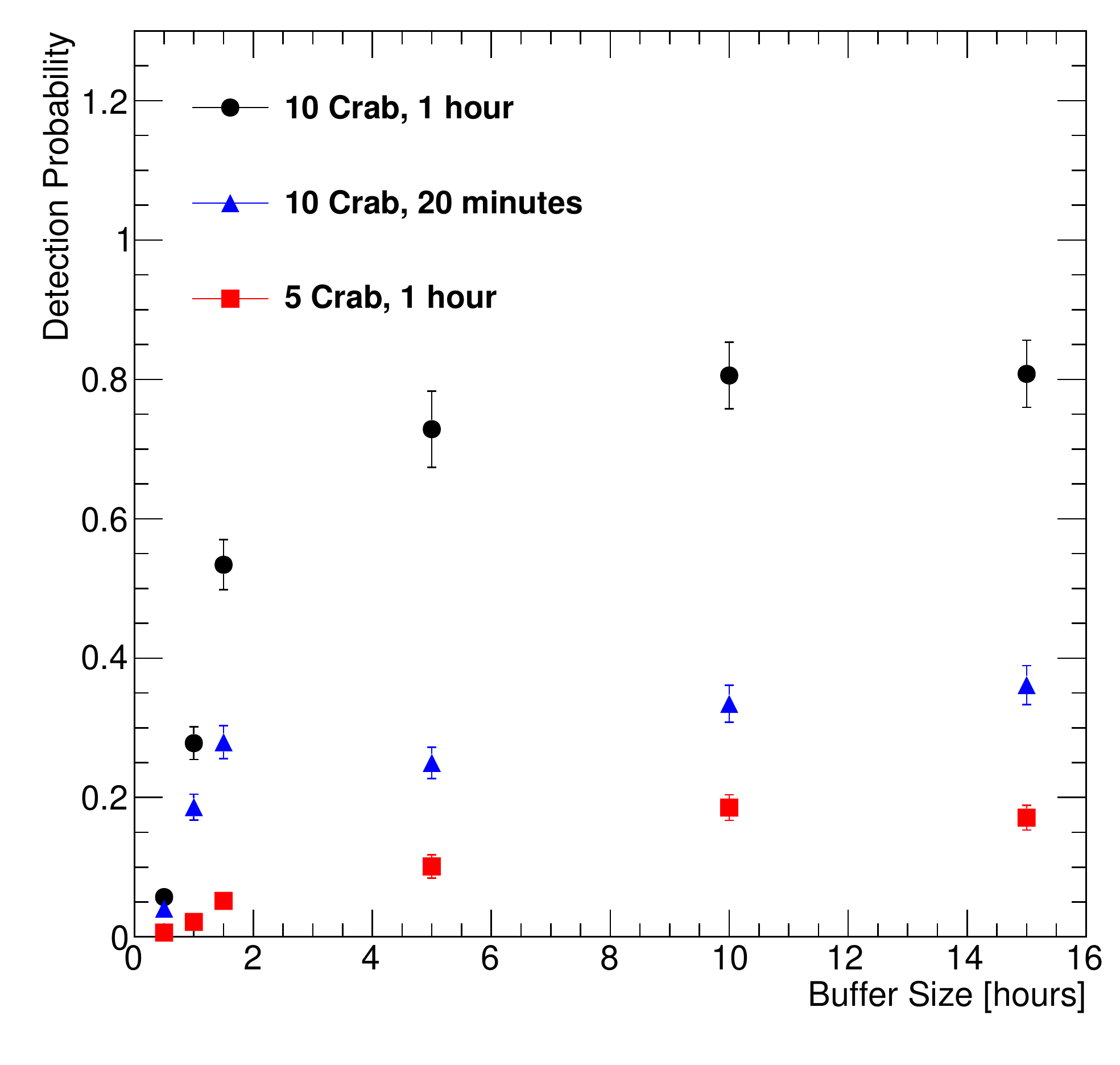}
  \caption{}
  \label{fig:monitor_sensitivity}
 \end{subfigure}
 \caption{
  \textit{(a):} Example of a simulated flare with a flux 10 times that from the Crab Nebula, with the rising edge identified by the flare monitor. The flare monitor was run with only 2 analysis bins to limit the number of plots.
  \textit{(b):} Probability for the optimal 3-bin flare monitor to detect flares of varying strength and duration, as a function of buffer size. We consider only those flares for which some part of the flare occurs when the event rate is at least half of its value at source culmination. The plot is made for a false positive rate of $10^{-5}$ triggers per minute per monitored source.
 }
 \label{fig:monitor_sensitivity_and_example}
\end{figure}

In order to characterize the sensitivity of the flare monitor, we inject into the observed data a number of simulated flares with spectra scaled to the spectrum of the Crab Nebula as measured by HAWC.
We then process the simulated results using the Bayesian block algorithm from Section~\ref{section:flares}, determining the number of simulated flares that are detected.
A sample simulated flare with a flux of 10 times that of the Crab Nebula appears in Figure~\ref{fig:monitor_example}.
For clarity, the figure depicts the results from running the flare monitor with 2 analysis bins instead of the optimal value of 3.
The flare monitor identifies the beginning of the flare as a significant change point, demonstrating its ability to detect flares.

Since the goal of the online flare monitor is to issue rapid alerts, we do not need to run the algorithm over the entire cumulative data set.
Instead, we run it over a buffer of configurable duration that contains only the most recent events.
Figure~\ref{fig:monitor_sensitivity} depicts the probability for the flare monitor to identify flares of varying strength and duration, as a function of buffer size.
In this case, we have run the flare monitor in the optimal case with 3 analysis bins.
As shown in the figure, in all cases the detection rate rises quickly at small buffer sizes, but then begins to level off, suggesting that buffer sizes larger than 10 hours have little effect on the sensitivity.
For a sufficiently large buffer, the probability to detect a flare that reaches 10 times the flux from the Crab Nebula for a period of 1 hour is approximately 80\%.
The figure also shows that the detection probability depends more strongly on the strength of the flare than its duration: a factor of 3 reduction in duration results in a detection probability of approximately 35\%, whereas dropping the flare strength by only a factor of 2 produces a detection probability of less than 20\%.

\section{Alert Expectations and Future Improvements}
\label{section:conclusions}

At present, the HAWC online flare monitor alerts are being tested internally by the HAWC Collaboration.
When the design of the system is finalized, alerts will be made available via email when significant flares are detected.
We expect to begin issuing these alerts soon.
Additionally, although the existing monitor already achieves sensitivity to flares that are known to occur, several improvements to the HAWC sensitivity are currently being investigated.
Among these are better gamma-hadron separation techniques and the development of a refined energy estimator to define the analysis bins.
When complete, these improvements will allow us to issue alerts for flares of lower strengths, as well as to more rapidly identify periods of extreme flaring activity.
Finally, we are developing two additional versions of the online flare monitor, one to search for Galactic transients, and another to conduct an all-sky search for transient VHE phenomena.
The main difference between the blazar flare monitor and these additional monitors will lie in setting the Bayesian prior, as rare events or large numbers of sources require a much lower false positive rate.

\section*{Acknowledgments}
\footnotesize{
We acknowledge the support from: the US National Science Foundation (NSF);
the US Department of Energy Office of High-Energy Physics;
the Laboratory Directed Research and Development (LDRD) program of
Los Alamos National Laboratory; Consejo Nacional de Ciencia y Tecnolog\'{\i}a (CONACyT),
Mexico (grants 260378, 55155, 105666, 122331, 132197, 167281, 167733);
Red de F\'{\i}sica de Altas Energ\'{\i}as, Mexico;
DGAPA-UNAM (grants IG100414-3, IN108713,  IN121309, IN115409, IN111315);
VIEP-BUAP (grant 161-EXC-2011);
the University of Wisconsin Alumni Research Foundation;
the Institute of Geophysics, Planetary Physics, and Signatures at Los Alamos National Laboratory;
the Luc Binette Foundation UNAM Postdoctoral Fellowship program.
}

\bibliography{icrc2015-0732-arxiv}

\end{document}